\documentclass[fleqn]{2023SCGE}
\usepackage[utf8]{inputenc}
\usepackage[numbers]{natbib}
\usepackage[switch]{lineno}
\let\WriteBookmarks\relax
\def\floatpagepagefraction{1}
\def\textpagefraction{.001}

\usepackage{xcolor} 
\usepackage{colortbl} 
\usepackage[normalem]{ulem}
\def\kms{$\rm{km\,s^{-1}}$}
\newcommand{\zcj}[1]{\textcolor{black}{#1}}
\usepackage{amssymb}
\usepackage{amsmath}
\usepackage{mathtools}
\usepackage{color}
\usepackage{enumitem}
\usepackage{textcomp}
\usepackage{threeparttable}
\usepackage{color}
\usepackage{epsfig}
\usepackage{color}
\usepackage{graphicx}
\usepackage{longtable}
\usepackage{soul, subfiles}

\usepackage{caption2}

\begin{document}
\let\WriteBookmarks\relax
\def\floatpagepagefraction{1}
\def\textpagefraction{.001}
\ensubject{subject}
\ArticleType{Article}
\Year{2023}
\Month{January}
\Vol{66}
\No{1}
\DOI{??}
\ArtNo{000000}
\ReceiveDate{??}
\AcceptDate{??}
  
\title {``Frog-eyes'' in Astronomy: Monitoring Binary Radial Velocity Variations Through A Pair of Narrow-Band Filters}

\author[1,2]{Chuanjie Zheng}{}
\author[2,1]{Yang Huang}{{yanghuang@ucas.ac.cn}}
\author[1,2,3,4]{Jifeng Liu}{{jfliu@nao.cas.cn}}
\author[1,2]{Hongrui Gu}{}
\author[1]{Hong Wu}{}
\author[1,2]{Youjun Lu}{}
\author[1,2]{\\Yongkang Sun}{}
\author[1]{Henggeng Han}{}
\author[1,4]{Song Wang}{}
\author[5]{Timothy C. Beers}{}
\author[2]{Kai Xiao}{}
\author[1,2]{\\Zhirui Li}{}
\author[1,2]{Boweng Zhang}{}
\author[1]{Yongna Mao}{}
\author[6]{Zhengyang Li}{}
\author[6]{Hangxin Ji}{}
\AuthorMark{Zheng C J, Huang Y, Liu J F}
\AuthorCitation{Zheng C J, Huang Y, Liu J F et al}

\address[1]{National Astronomical Observatories, Chinese Academy of Sciences, Beijing 100101, China}
\address[2]{School of Astronomy and Space Science, University of Chinese Academy of Sciences, Beijing 100049, China}
\address[3]{New Cornerstone Science Laboratory, National Astronomical Observatories, Chinese Academy of Sciences, Beijing 100101, China}
\address[4]{Institute for Frontiers in Astronomy and Astrophysics, Beijing Normal University, Beijing 102206, China}
\address[5]{Department of Physics and Astronomy, \\University of Notre Dame and Joint Institute for Nuclear Astrophysics -- Center for the Evolution of the Elements (JINA-CEE), Notre Dame, IN 46556, USA}
\address[6]{Nanjing Institute of Astronomical Optics and Technology, Chinese Academy of Sciences, Nanjing 210042, China}

\abstract{
Spectroscopic observations are a crucial step in driving major discoveries in the era of time-domain surveys. However, the pace of current spectroscopic surveys is increasingly unable to meet the demands of rapidly advancing large-scale time-domain surveys.
To address this issue, we propose the ``Frog-eyes'' system, which employs a pair of narrow-band filters: one positioned near a strong absorption line to capture signals from Doppler shifts, and the other placed on the adjacent continuum to monitor intrinsic variations.
The combination of observations from the two filters enables  the extraction of radial velocity (RV) curves from a large sample of binary stars, and is particularly efficient for single-lined binaries (SB1), using photometric techniques.
Comprehensive mock simulations on SB1 demonstrate that the binary orbital parameters can be precisely measured from the extracted RV curves for binary systems where the primary star has an effective temperature greater than 6000~K.
With a typical ground-based photometric precision of approximately 0.3\%, the uncertainties in the derived semi-amplitude $K$ and eccentricity $e$ are less than 10\% and 0.1, respectively, for binary systems with $K \ge 30$~\kms.
These encouraging results are further validated by real observations of the hot subdwarf-white dwarf binary system HD 265435, using a non-specialized ``Frog-eyes'' system installed on the Chinese 2.16m telescope.
Once this system is properly installed on large-field-of-view survey telescopes, the rate of acquiring RV curves for binaries will \zcj{approach} their detection rate in leading time-domain photometric surveys.}



\maketitle

\begin{multicols}{2}
\section{Introduction}
Large-scale astronomical surveys play a pivotal role in modern astrophysical studies, unveiling new discoveries and fundamentally reshaping our understanding of the origin and evolution of the Universe.

Fast technological advances are changing the ways we view the sky, from snapshot captures, as demonstrated in
\Authorfootnote
\noindent
projects like the Sloan Digital Sky Survey \citep[SDSS;][]{SDSS} and Pan-STARRS1 \citep[PS1;][]{ps1}, to continuous monitoring through surveys such as $Gaia$ \citep{gaia0}, the Zwicky Transient Facility \citep[ZTF;][]{ZTF} and the Large Synoptic Survey Telescope \citep[LSST;][]{LSST}, which repeatedly observe the same sky hundreds to thousands of times.
In the future, we will have telescope arrays like SiTian \citep{sitian} that can scan the entire sky at a cadence of every 30 minutes.
These developments open the possibility for probing the dynamic Universe either from rapid flux variations or fast celestial movements across the sky.

\begin{figure*}
    \centering
    \includegraphics[scale=0.7, angle=0]{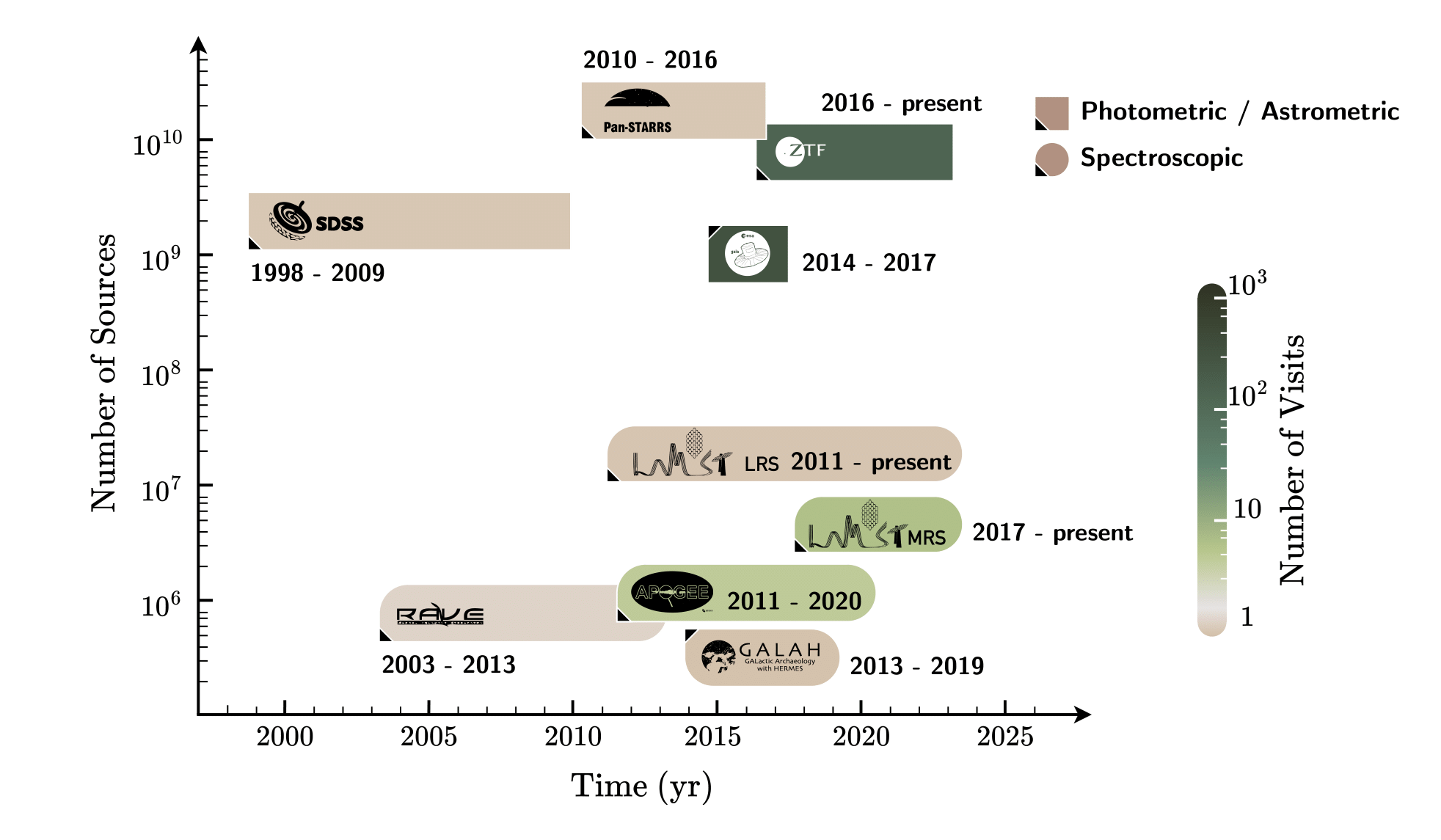}
    \caption{\textbf{Data Volume and average visits of several well-known sky survey projects.}  The average number of visits are represented by the color with the colorbar shown on the right side of the plot. Black right-angle triangles mark the beginning time and number of sources for different projects. }
    \label{fig:volume}
\end{figure*}

As another important pathway to understand the Universe, spectroscopic surveys have also undergone tremendous development over the past half-century. 
From hundreds of thousands to millions of spectra obtained from projects like the Radial Velocity Experiment \citep[RAVE;][]{RAVE}, SDSS, the Apache Point Observatory Galactic Evolution Experiment \citep[APOGEE;][]{Apogee}, and the Galactic Archaeology with HERMES \citep[GALAH;][]{galah}, to the mega spectroscopic surveys like the Large Sky Area Multi-Object Fiber Spectroscopic Telescope \citep[LAMOST;][]{LAMOST1,LAMOST2,LAMOST3}, which have collected tens of millions of spectra, these surveys have deeply influenced our understanding of stars, the Milky Way, and even the entire Universe.
However, even given this progress, the number of spectroscopic targets lags far behind those of the photometric or astrometric surveys mentioned above by at least 2-3 orders of magnitude (see Figure~\ref{fig:volume}).
More seriously, the number of target sources with repeated spectroscopic observations is only a few million, with an average number of visits of around five per source, which is insufficient for measuring the radial velocity (RV) amplitude of a binary star.
This lag in spectroscopic observation capabilities, compared to the advances in photometry and astrometry, has become the primary bottleneck of numerous fields, notably within the burgeoning field of time-domain astronomy.
Even when combined, all current and upcoming spectroscopic facilities can follow up only a small fraction of the transient alerts detected by ZTF, let alone the future super transient factories LSST or SiTian.

To partially alleviate this issue, we propose an astronomical ``Frog-eyes'' system that can photometrically measure the RV curves of binary systems using a pair of narrow-band filters, with one close to a strong absorption line (hereafter the $A$-band) and the other sitting on the nearby continuum (hereafter the $C$-band).
The $A$-band is primarily designed to capture the flux variations caused by the Doppler shifts of the primary in a binary system, while the $C$-band can record any flux variations, whether from modulations of the companion (such as eclipsing or tidal distortion) or from the primary itself (such as rotation or pulsation).
In this regard, the RV \zcj{variation} of a binary system, whose fraction is as high as 50\% in the Galaxy, can be accurately derived by comparing observations obtained from the two filters.
\zcj{Single-lined spectroscopic binaries (SB1) are particularly well-suited for this method, as their $A$-band flux is unaffected by the anti-phase motion of a secondary component. This allows precise orbital parameter measurements for SB1 systems, even with the photometric precision easily achievable using ground-based telescopes.}
When installed on a large field-of-view telescope, this \zcj{paired-filter} system could acquire the RV curves of binary systems at a rate comparable to their detection in leading time-domain surveys like ZTF.
This effort, along with the rich imaging and astrometric data, will help us better understand the physical processes at play in the dynamic Universe.
  
This paper is organized as follows. In Section 2, the detailed conceptual design of this narrow-band \zcj{paired-filter} system is introduced. Using mock data, we then check the performances of this filter system in deriving key orbital parameters (e.g., semi-amplitude of RV curve $K$ and eccentricity $e$) for different \zcj{SB1} in Section~3.
In Section 4, we introduce the first successful attempt to derive the RV curve of a close binary system using real observations from a narrow-band \zcj{paired-filter} system installed on the Chinese 2.16 m telescope.
\zcj{Further discussion on the current caveats and challenges of the ``Frog-eyes'' design, along with our efforts to address and resolve them, is presented in Section~5.}
Finally, a summary is given in Section~6.

\section{Conceptual Design}
\label{sec:concept}
The key to this methodology is the photometric measurement of the RV curve for a \zcj{SB1 system}.
Historically, there has been a few attempts \citep{BEAM_planet, BEAM_Detect_space, BEER0} that mainly relied on the beaming effect, where the binary's brightness is modulated by the reflex motions of stars in binary systems.
However, such modulation is tiny, typically on the order of $BK/c \sim 10^{-4}$ to $10^{-3}$, where $c$ is the speed of light and $B$ refers to the beaming factor. 
Monitoring such tiny variations in brightness requires ultra-high photometric precision, achievable only through space-based missions like CoRoT \citep{COROT,corot2}, {\it Kepler} \citep{kepler}, and TESS \citep{TESS}.
Despite significant advancements in space missions over the past decades, less than 100 RV curves of binary systems have been measured this way \citep{BEER2}.

\begin{figure*}
    \centering
    \includegraphics[width=\textwidth]{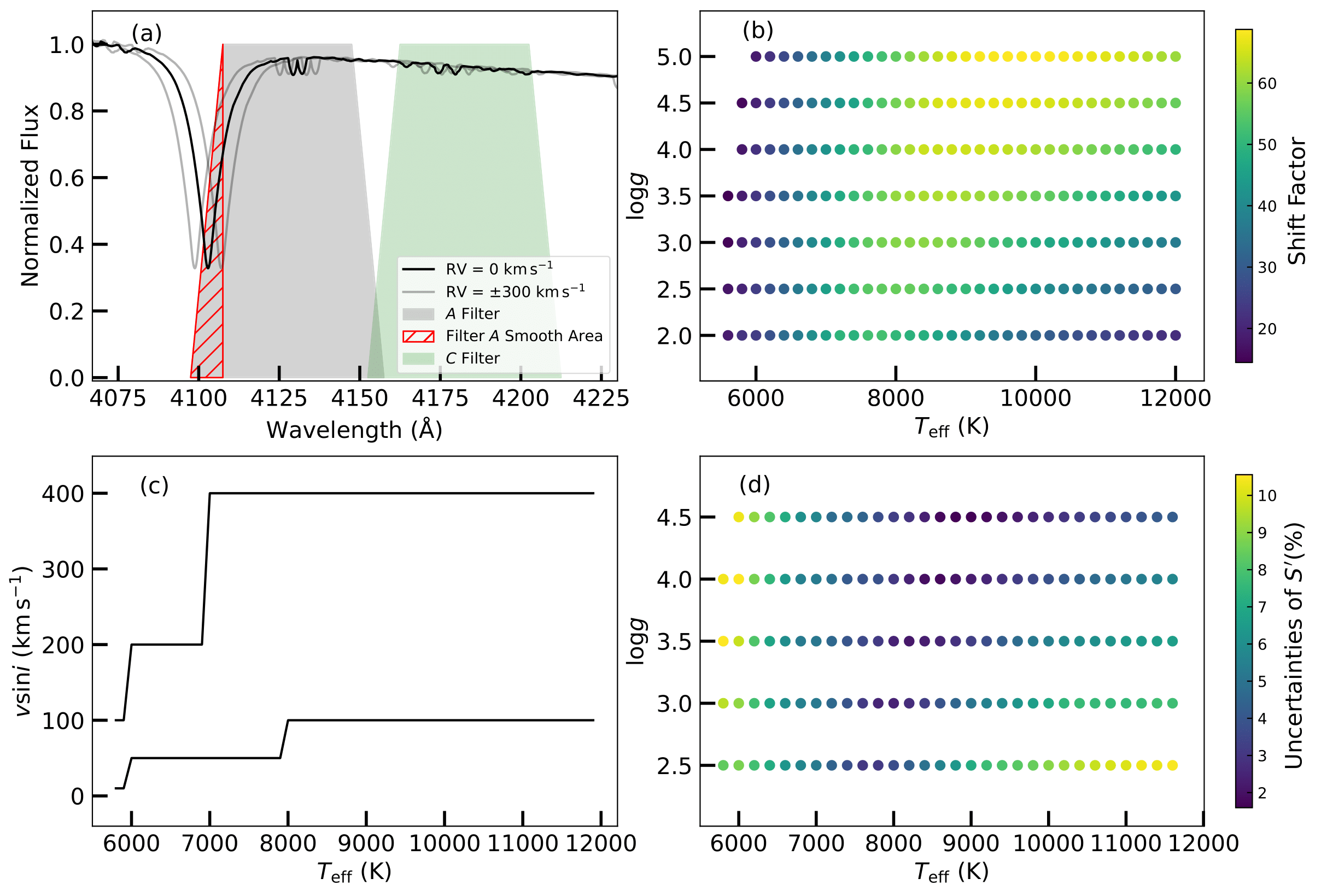}
    \caption{\textbf{An example design \zcj{for} the $A$- and $C$-band and the shift factor of the $A$-filter.} \textbf{(a)} Transmission curves are shown for the $A$- (gray shaded region) and $C$-band (green shaded region). The atmospheric parameters of the background spectra are: $T\rm_{eff}=12,000$~K, log~$g$ = 3.5, $v$sin$i$ = 100 \kms, and [Fe/H] = 0. The left edge of the $A$-band is 4097.5 \AA\, and the right edge is 4157.5 \AA\,. The width of the short side of both left and right right-angle triangular is 10 \AA. The $C$-band has a similar profile to the $A$-band, but is red-shifted by 55 \AA\,. \textbf{(b)} Shift factors of the $A$-band within its applicable parameter ranges ($D_2 < 10$\%) for solar-metallicity systems with $v$sin$i$ = 100 \kms.\textbf{(c)} Upper and lower limits of $v \sin i$ across different effective temperature ranges. The lower $T\rm_{eff}$ range ($\le 5700$\,K) is not shown because the $A$-band is not applicable there. \textbf{(d)} Uncertainties of $S'$ near solar metallicity when considering uncertainties in atmospheric parameters and $v$sin$i$. The parameter ranges are narrower than the applicable $A$-band ranges shown in panel (b), where the uncertainties are difficult to calculate due to edge effects.}
    \label{fig:example}
\end{figure*}

We therefore propose a new technique: placing a narrow-band filter near the strong absorption lines of \zcj{stars, such as the Balmer series} (referred to as the $A$-band).
The binary motion would cause the absorption line of the visible star to frequently move in and out of the $A$-band, resulting in notable flux variations.
As shown in Figure~\ref{fig:example}(a), by placing a narrow-band filter with an effective width of approximately 50~\AA~ (see discussions of filter design in Supplementary Sections A and B) positioned near the H$\delta$ line, which can be easily constructed with current technology, the $A$-band photometric amplitude induced by the Doppler shift can reach variations of one-tenth of a percent to a few percent.
This is an order of magnitude higher than the effect caused by beaming, making it easily detectable by current ground-based telescopes.
The significant modulation on flux is due to the steep slope of the absorption line's wing. In the context of Doppler beaming, such a steep slope implies a very large \zcj{absolute value of }spectral index. Consequently, the `pseudo beaming factor' (hereafter shift factor) for this narrow $A$-band is expected to be high (an order of magnitude larger than the usual beaming factor), as it is approximately proportional to the spectral index.

Quantitatively, similar to the beaming effect, the Doppler shift flux in the $A$-band can be expressed as:
\begin{equation}
    F_{A}(t) = \eta(t) F_{0,A}(t)\left(1-S\,\frac{v_{r}^{\rm Earth}}{c}\right)\text{\,,}
    \label{eq:ha2}
\end{equation}
where $F_{A}(t)$ denotes the observed time-series flux in the $A$-band for a target, $F_{0,A}(t)$ represents the intrinsic time-series flux without the modulation from the Doppler shift. $\eta$ refers to the overall efficiency in this wavelength region of the $A$-band, including but not limited to Galactic reddening, atmospheric extinction, and the telescope total efficiency. 
$S$ is the $A$-band shift factor, which can be calculated for any type of star using the method developed in \citep{Z24} (hereafter Z24).
$v_{r}^{\rm Earth}$ refers to the radial velocity in the Earth's frame, which can be expressed as: 
$v_{r}^{\rm Earth} = v_{r} + v_{\oplus}$. Here, $v_{r}$ represents the radial velocity in the heliocentric frame, and $v_{\oplus}$ represents the component of Earth's radial velocity relative to the Sun in the direction of the target.

Although the $S$ factor could be an order of magnitude larger than the beaming factor, there are still two challenges in measuring the binary RV curve from this specially designed $A$-band.
First, how to achieve a constant shift factor $S$ when converting flux into $v_{r}$. This is necessary because the RV curve cannot be determined under a RV-dependent factor $S$ due to the lack of knowledge about the systemic radial velocity $v_{\gamma}$ of the binary system.
Secondly, how to separate between flux modulations caused by the Doppler shift and other effects reflected in $F_{0,A}(t)$, such as eclipsing or tidal distortion from the companion, as well as rotation of the primary itself. \zcj{We note that other factors, such as stellar spots or accretion, can intermittently affect the spectral line profile, leading to orbital-phase-dependent variations in the measured RV curve. This could limit the precision of orbital parameter measurements. However, such events are relatively rare, and \zcj{are} discussed in detail in Section~5.}

For the first issue, the key is to design a specific shape at the edge of the $A$-band that allows for a smooth reception of flux variations caused by the Doppler shift.
Here we adopt a trapezoidal profile design for the $A$-band (see Figure~\ref{fig:example}(a)).
The key `smooth area' is the right-angled triangular region (highlighted by red diagonal lines) at the left edge of the $A$-band. This region is designed to approximately linearly modulate the flux as the absorption line moves in or out.
Based on the detailed computations in Section \ref{sec:range}, the $S$ factor of the $A$-band in such a design can be considered a constant value for many types of stars.
We note that this trapezoidal shape is just one approach to achieve linear responses. We thus encourage the exploration of alternative designs that may offer improved performance.

For the second issue, we introduce an additional filter with a transmission curve \zcj{shape identical to that of the $A$-band,} but with the central wavelength red-shifted by approximately $55$~\AA~ (see Figure~\ref{fig:example}(a)), where no significant strong absorption lines are present.
This filter is adopted to monitor flux variations from effects other than \zcj{the} Doppler shift (hereafter referred to as the $C$-filter).
Due to the minimal difference in central wavelength between the $A$- and $C$-bands, their flux ratio can be approximated as a constant value: $F_{0,A}=m\,F_{0,C}$, where $m$ is expected to be close to one in principle.
Considering the usual beaming effect, the observed $C$-band flux can be expressed as:
\begin{equation}
    F_{C}(t) = \xi(t) F_{0,C}(t)\left(1-B\,\frac{v_{r}^{\rm Earth}}{c}\right)\text{\,,}
    \label{eq:ha3}
\end{equation}
where $\xi (t)$ is approximately equal to $\eta (t)$, given the minor difference in central wavelength between the $A$- and $C$-bands. $B$ is the beaming factor, which can be easily calculated using the method developed by Z24.
We note that this design just serves as an example. For an alternative design that is slightly less effective in monitoring intrinsic flux variations but optimized for amplifying RV-modulated signals, please refer to Supplementary Section B.

In realistic observations, the light curve can be constructed using the differential photometry method that \zcj{adopts} a zero-point at a chosen time $T_A$/$T_C$ for the $A$-/$C$-bands, respectively.
The combination of the two light curves obtained from \zcj{the}
$A$- and $C$-bands yields:
\begin{equation}
    \frac{F_A(t)}{F_C(t)} = m\frac{\eta(T_A)}{\xi(T_C)}\frac{1- S\frac{v_{r}^{\rm Earth}}{c}}{1- B\frac{v_{r}^{\rm Earth}}{c}} \text{.}
    \label{eq:divide_0}
\end{equation}
Given $B\frac{v_{r}^{\rm Earth}}{c} \sim 0$, Equation\,\ref{eq:divide_0} can be further expressed using a small-quantity approximation:
\begin{equation}
    \frac{F_A(t)}{F_C(t)} \approx m' \left( 1- S'\frac{v_r^{\rm Earth}}{c}\right)\text{,}
    \label{eq:divide}
\end{equation}
where $m'$ is $m\frac{\eta(T_A)}{\xi(T_C)}$, which should also be a constant. $S'$ represents $S - B$.
For a \zcj{SB1} system, $v_r = v_{\gamma} + \delta v$, where $v_{\gamma}$ and $\delta v$ denote the systemic velocity and radial component of orbital motions of \zcj{the visible star}.
Thus, Equation\,\ref{eq:divide} can be further written as:
\begin{equation}
    \frac{F_A(t)}{F_C(t)} \approx m' \left( 1- S'\frac{v_{\gamma}+v_\oplus}{c}\right)- m'S'\frac{\delta v}{c}\text{.}
    \label{eq:divide_f}
\end{equation}
\label{sec:V0}
In this equation, both parameters $S$ and $B$ can be determined from either synthetic or real spectra. The value of $v_\oplus$ can be accurately measured using known coordinates of stars, the observatory's location, and the observation epoch. Although $v_{\gamma}$ is unknown, the term $S' \frac{v_{\gamma}}{c}$ can be neglected as it is generally smaller than 2\% (most Galactic field stars have absolute RV values within $100$ km s$^{-1}$, as shown in Supplementary Figure S4). Therefore, the RV curve of a binary system, $\delta v(t)$, can be accurately measured using the combined light curves obtained from the $A$- and $C$-bands. 

In summary, just as a frog’s eyes are adept at capturing moving objects, this pair of narrow-band filters is highly effective at detecting line shifts, thus earning the name ``Frog-eyes'' system. In particular, for systems exhibiting single-lined periodic motions, such as SB1 binaries and pulsating stars, their RV curves can be fully modeled. As highlighted in this study and discussed in the next section, orbital parameters, such as the RV semi-amplitude $K$ and eccentricity $e$, can be precisely derived from the RV curve measured by our ``Frog-eyes'' system for SB1 systems.
For systems exhibiting multi-lined motions, aperiodic motions, or even line profile changes (such as those caused by stellar spots or accretion, as mentioned above), the ``Frog-eyes'' system can also readily detect these phenomena. When combined with broadband light curves and follow-up spectroscopy, the resulting data can provide valuable insights into the nature of these intriguing systems. However, this is beyond the scope of the current study and will be explored in future work. This study will focus on the ability of the ``Frog-eyes'' system to measure the orbital parameters of SB1 systems, which are discussed in detail in Sections 3 and 4 through both mock simulations and test observations.


\section{Performance of an Example Design}
In this section, we conduct a comprehensive set of mock data tests to evaluate the ability of the narrow-band filter pair in \zcj{photometrically} measuring the binary orbital parameters (i.e., $K$ and $e$) from \zcj{the light curves} \zcj{of a SB1 system}. To do so, we first present the detailed parameters of an example design of this narrow-band filter pair.
We note that this is only one possible design, taking into account the applicable range, magnitude, and linearity of the shift factor, as well as limiting magnitude (see Supplementary Sections A and B). Secondly, we evaluate the uncertainties in binary orbital parameters using light curves from this narrow-band filter pair (see Equation~\ref{eq:divide_f}), considering typical photometric errors from current surveys and a wide range of binary orbital parameters.

\subsection{Filter Design and Its Applicable Range}
\label{sec:range}
Designing the $A$-filter involves two key steps: 1) defining the main body, controlled by two parameters -- central wavelength and effective width; and 2) defining the edge shape, which serves as the `smooth area', and is crucial for detecting shifts in the stellar absorption line.
For the main body, we first need to determine its central wavelength, positioning it close to one of the strong stellar absorption lines.
In this study, \zcj{the} H$\delta$ ($4102$~\AA) line is chosen due to its deep absorption and steep line wing.
Secondly, to achieve a high shift factor, the filter’s effective width should be as narrow as possible. However, to reach a deep limiting magnitude, a wider width is necessary.
Here, we choose an effective width of 50 \AA\ to balance \zcj{these} considerations.

As discussed in Section~2, the shape of the edge is crucial for linearly converting Doppler shifts into flux modulations.
Its position is also a crucial parameter for determining the magnitude of the shift factor.
After several tries, the right-angled triangular shape with the shorter side of 10~\AA~ is finally chosen in this study.
In summary, this design, as shown in Figure \ref{fig:example}\,(a),  can simultaneously meet our requirements for the magnitude of the shift factor, its applicability range, limiting magnitude, and  manufacturing difficulty (see Supplementary Section A for detail discussions).
However, we note that it \zcj{could} be continuously improved, as the exploration of other Balmer lines (such as H$\gamma$) and edge shapes have not yet been tested.
The $C$-filter has the same profile as the $A$-filter, but is red-shifted by 55~\AA~ in wavelength.

We adopt the technique developed by Z24 to precisely determine the $A$-band shift factors of different types of stars across a RV range from $-300$\,\kms\ to $+300$\,\kms\,(with a step of 2.5 \kms) using synthetic spectra from the \texttt{PHOENIX} library \citep{PHOENIX}.
The $D_2$ index\footnote{This index tracks the fraction of data points deviating from the linear fit by $\frac{\delta K}{K}F_{\rm Amp}$, where $\delta K  = 2\%K$ and $F_{\rm Amp}$ represent the difference between the flux peak and valley, respectively.} proposed by Z24 is used once more in this study to evaluate the linearity of the resulting shift factor. We propose that the flux variation responds linearly to the Doppler shift if the $D_2$ index of the shift factor is less than 10\%.
During the computation, the adopted parameter coverage of the \texttt{PHOENIX} library is as follows: effective temperature $T_{\rm eff}$ ranging from 3000\,K to 12,000\,K, surface gravity log~$g$ ranging from 2.0 to 5.0, and metallicity [Fe/H] ranging from $-3.0$ to $+1.0$. The average step sizes are 200\,K for $T_{\rm eff}$, 0.5 dex for log~$g$, and 0.5 dex for [Fe/H], respectively.
In addition to the default version of the \texttt{PHOENIX} library, we also compute shift factors for spectra with varying rotational velocities, by convolving synthetic spectra 
with different values\footnote{The grid for $v \sin i$ includes values of 15, 30, 50, 70, 100, 150, 200, 250, 300, and 400~\kms.} of $v \sin i$ using the python procedure \texttt{LASPEC} \citep{laspec1,laspec2}.

The results demonstrate that the $A$-band has an exceptional ability to linearly amplify Doppler shift on flux variations across a broad range of stellar types. 
For instance, considering solar-metallicity stars with $v$sin$i$ = 100~\kms, the shift factor remains relatively constant with $D_2 < 10\%$ for effective temperatures between $6000$ and \zcj{$12,000$~K}. Within this range, the median shift factor $S$ is notably high, approximately 50, which is at least an order of magnitude greater than typical beaming factors.
This applicable range and the corresponding shift factors are illustrated in Figure\,\ref{fig:example}\,(b).
The results reveal a distinct `$\Lambda$' pattern in the plot of shift factors versus $T_{\rm eff}$, where shift factors initially increase and then decrease with \zcj{increasing} $T_{\rm eff}$. This pattern inherits the evolution of the equivalent width of the H$\delta$ line with $T_{\rm eff}$.
For stars with $T_{\rm eff}$ below 6000~K, the shift factors exhibit non-linearity, becoming RV-dependent. This behavior arises from the weakening of the H$\delta$ line for stars cooler than 6000~K and the emergence of molecular lines, especially for stars with $T_{\rm eff} < 4000$~K.
The upper applicable $T_{\rm eff}$ limit is due to the limitation of the \texttt{PHOENIX} library.
In principle, the $A$-band shift factors can remain constant for $T_{\rm eff} \ge 12,000$~K. In the near future, we plan to calculate shift factors for this higher $T_{\rm eff}$ range using other synthetic libraries designed for hotter stars ($T_{\rm eff} \ge 10,000$~K).

In addition, similar procedures, including computations of the beaming factor and the $D_2$ index, are also performed for the $C$-band. Within the applicable linear-approximation range of the $A$-band, the median value of beaming factor $B$ is found to be 3. Significant non-linearity ($D_2\ge$10\%) is detected for stars with $T_{\rm eff} < 8000$~K. However, the non-linearity's effect on the flux, which is less than 0.1\%, is minor for ground-based telescope detections, and can therefore be neglected in Equation \ref{eq:divide_f}.

\subsection{Ability to Extract Orbital Parameters}
In this subsection, we present detailed simulations to demonstrate the capabilities of the $A$- and $C$-bands in extracting orbital parameters.
As noted earlier, the simulations are performed for SB1 systems.
During the simulation, two main issues are considered: 1) the precision of the light curves, and 2) the errors in modeling the shift and beaming factors for a given target.
This simulation primarily evaluates the measured uncertainties for varying values of $K$ and $e$ \zcj{ in different types of SB1 system,} regardless of the specific binary populations. To illustrate this, we use a star with atmospheric parameters of $T_{\rm eff}=10,800$~K, $\log{g}=3.5$, [Fe/H]$ = 0$, and $v\sin{i} = 100$~\kms in the subsequent calculations. \zcj{This system is selected, as its shift factor is approximately 50, a typical value for} stars with $D_2 < 10\%$ (see Figure \ref{fig:example}\,(b)). 

\subsubsection{Mock Observations}
\label{sec:mock}
The light curve simulations include two parts: 1) generation of theoretical light curves under a grid of binary orbital parameters \zcj{for four different sub-types of SB1}, and  2) modeleding observed light curves with a typical ground photometry precision and a given survey strategy. 

\zcj{First, to generate theoretical light curves for the $A$-band and $C$-band, we focus on two key factors that shape their curves: the signal from the line shifts, which predominantly affects the $A$-band, and the background from the star's intrinsic flux and its variations, which primarily influences the $C$-band while also contributing to the $A$-band.}

\zcj{1) The signal refers to the RV-modulated flux variation originating from line shifts and Doppler beaming. It is exactly characterized by two key features: the signal amplitude (determined by $K$) and the signal morphology (determined by eccentricity $e$ and the argument of periastron $\Omega$, when the light curve is folded into phase space). To achieve a higher signal-to-noise ratio, we binned the light curves in phase space. Notably, systems with identical $K$, $e$, and $\Omega$, but different periods. exhibit no differences in phase space. Consequently, the choice of orbital period ($P$) does not affect the simulation results.  Therefore, $P$ is fixed to a specific value (i.e., $P$ = 0.9 d), while $K$, $e$, and $\Omega$ are sampled to cover nearly the entire parameter space (see details in Table \ref{tab:params}), regardless of whether the parameter combinations are physically realistic.}
\zcj{In addition, }$v_\oplus$ is a well-understood parameter and was set to 0~\kms\ for simplicity. \zcj{The inclination angle $i$ was fixed to 70 degrees for the non-eclipsing cases and to 90 degrees for the eclipsing cases, as it does not affect the signal for a given $K$.}
\zcj{Then, t}he Python module \texttt{twobody}\footnote{https://github.com/adrn/TwoBody} was adopted to predict RV curves using \zcj{the} above grid.

\zcj{2) The background refers to the intrinsic flux  and its variations. For SB1 systems, these variations are primarily caused by eclipsing and ellipsoidal modulations. Based on these mechanisms, the types of binaries can be categorized as eclipsing binaries (wide, nearly edge-on binaries), ellipsoidal binaries (close, non-edge-on binaries), eclipsing+ellipsoidal binaries (EE; close, edge-on binaries), and non-EE binaries (wide, non-edge-on binaries). To ensure robust results in the simulation, we set the eclipse depth and ellipsoidal amplitude to relatively high values of 50\% and 10\%, respectively. Such levels are only observed in some extreme binary systems, and our obtained results can be considered as a conservative lower limit for performance of the ``Frog-eyes'' system. Additionally, as discussed in Section 2, the systemic velocity $V_\gamma$ has a minor effect on the measurement of intrinsic flux. To further ensure robustness, $V_\gamma$ was also fixed at a relatively large value of 40~\kms\ (1$\sigma$ value for all systems, see Supplementary Figure S4).}

Theoretical light curves are then generated by incorporating the shift and beaming effects as described by Equations \ref{eq:ha2} and \ref{eq:ha3} (the total efficiency at the two bands, i.e., $\eta$ and $\xi$, are set to a constant value of 100\% for simplicity in the calculations) \zcj{for four different types of SB1.} 

Secondly, to simulate observed light curves, we use typical photometric uncertainties and a straightforward observation strategy, as described below. The photometric precision for each individual epoch is set to 1\%, modeled as a Gaussian distribution. Additionally, a zero-point correction uncertainty of 0.3\% is applied to each exposure, which is considered to be the highest accuracy achievable for ground-based telescopes.
For the observations, we adopt a straightforward survey
\begin{table}[H]
    \footnotesize
    \centering
    \caption{Grid of orbital parameters.}
    \begin{tabular}{lll}
    \hline
        Parameter&unit & Value \\
        \hline
        $\rm log_{10}($$K)$ &$\rm{km\,s^{-1}}$& 0.6 to 2.6 step 0.1\\\hline
        Eccentricity $e$ & n.a.&0, 0.05, 0.1, 0.2, 0.5, 0.7\\\hline
        Argument of periastron $\Omega$ &degree& 0, 45, 90, 120, 150 \\\hline
        Period & day & 0.9\\\hline
        $v_{\gamma}$ & \kms & 40\\\hline
        Inclination angle $i$ & degree & 90 for eclipsing, 70 for non-eclipsing\\
        \hline
    \end{tabular}
    \label{tab:params}
\end{table}
\noindent strategy: each field will be monitored with a 1-minute cadence for three hours per night, over a period of weeks to months.
The strategy will consider observations from different stations and the detailed timing of the three-hour exposure each night to ensure that light curves cover as much of the phase as possible for binaries with varying periods.
The monitoring time interval is also considered in the strategy to ensure a sufficient number of visits in each phase bin (see the later discussion), thereby reducing random uncertainty to the best accuracy, i.e., 0.3\%.
As part of the SiTian project \cite{sitian}, it is assumed that the $A$- and $C$-filters are installed in two separate telescopes equipped with identical optical systems and cameras. This setup allows for the simultaneous collections of light curves in both bands. 
For each grid, 100 light curves are simulated separately for the $A$-band and the $C$-band. 
In this way, a total of \zcj{504,000 mock light curves ($630$ grids $\times$ $100$ stars $\times$ 2 bands $\times$ 4 types)} were generated for subsequent analysis.

With the simulated observational light curves, we can now evaluate the typical uncertainties in the extracted orbital parameters.
The entire procedure is outlined in detail below: 

1)\,Determine the orbital period. At this step, we apply the Lomb-Scargle method directly to the shift curves (i.e., $\frac{F_A}{F_C}$) to prevent potential ambiguities from extraneous periodic signals such as \zcj{eclipses}.
The frequency corresponding to the highest power is then selected as the period.

2)\,Fold the shift curves into a phase diagram. 
The entire phase is divided into 100 bins, and the median value and its error are calculated for each bin.
After binning, photometric uncertainties are reduced to 0.3\% in most cases (for example, see Supplementary Figure S3).

3)\,Model the binned shift curves by Equation \ref{eq:divide_f}. The obtained shift curves are modeled by Equation \ref{eq:divide_f} using {\tt The Joker} \citep{thejoker}, which is modified slightly.
During the modeling procedure, $v_\gamma$ is set to 0\,\kms. 
The best fit yields the orbital parameters (including semi-amplitude $K$, eccentricity $e$, and argument of periastron $\Omega$), as well as the constant value $m'$, which is related to the observational system and specified weather conditions.

4)\,Compute the uncertainties. 
As mentioned above, we generate 100 stars for each parameter grid.
They are treated as a Monte Carlo (MC) simulation for each grid. The difference between the true value and the median of the distribution yielded by the MC simulation is considered as the systematic error, while the scatter of the distribution represents the random uncertainty. 

\begin{figure*}
    \centering
    \includegraphics[width=0.95\textwidth]{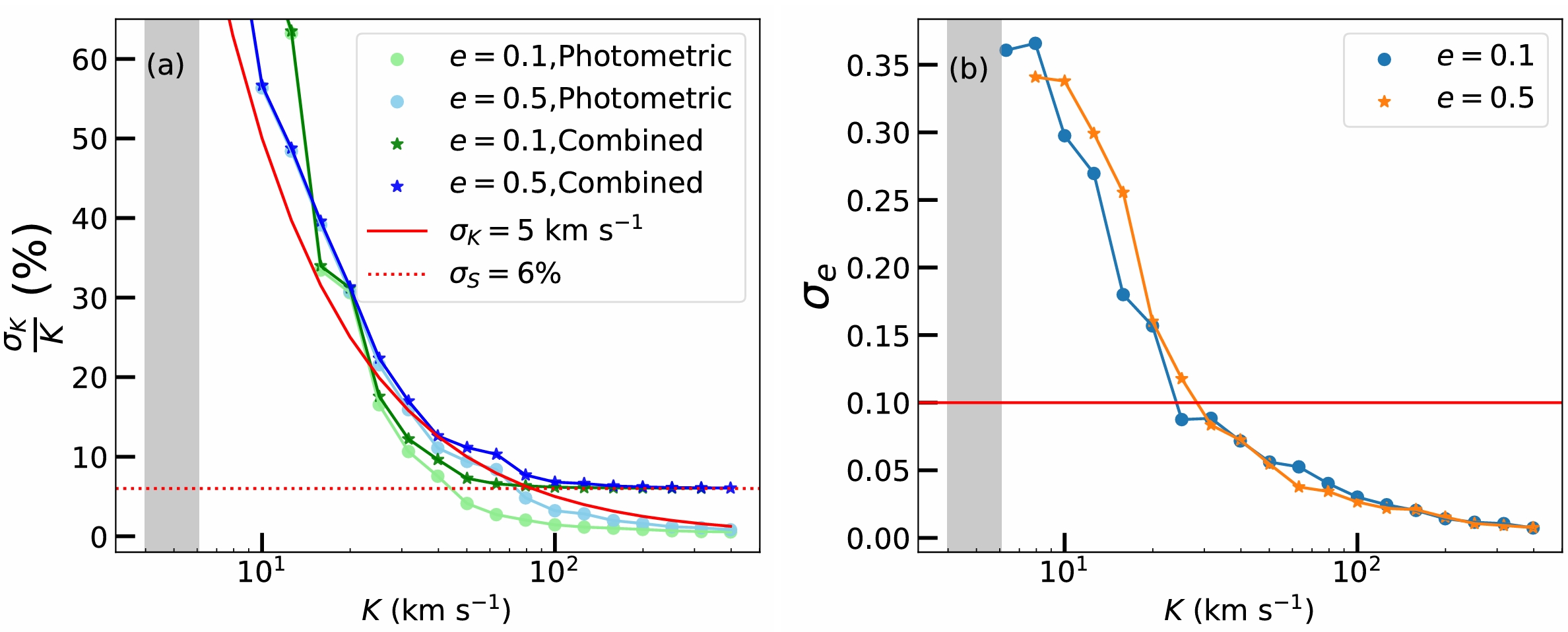}
    \caption{\textbf{Uncertainties in orbital parameter measurements for \zcj{a} ellipsoidal plus eclipsing SB1 system.} The target system is assumed to be a star with $T_{\rm eff} = 10,800$\,K, $\log{g} = 3.5$, [Fe/H] = 0, and $v\sin{i} = 100$ \kms. The arguments of periastron ($\Omega$) are randomly selected to be 90 degrees due to their negligible influence. The shift factor of the $A$-band and the beaming factor of the $C$-band are 50.2 and 2.13, respectively. The shift factor uncertainty is $6\%$. The gray areas in both panels denote where the periodic modulation signal is too noisy to detect \zcj{with a} Lomb-Scargle diagram.  \textbf{(a)} Percentage uncertainties of $K$ for different $K$ values. The dot symbols denote uncertainties arising from photometric observations. The red dashed line denotes uncertainties arising from shift factor uncertainties. The star symbols show the results from error propagation of the two components. \textbf{(b)} Uncertainties of eccentricity $e$ at different $K$ values.}
    \label{fig:vary}
\end{figure*}

Overall, the uncertainties in the extracted values of $K$ and $e$ exhibit a strong dependence on the magnitude of $K$, with a moderate influence from the eccentricity (see Figure \ref{fig:vary}).
\zcj{Additionally, they exhibit negligible dependence on the types of binaries (see Figure \ref{fig:vary} for the most complex case, the ellipsoidal plus eclipsing binary, and Supplementary Figures S6 and S7 for the other three types of binaries).
This is easily understood, as the type of binary only affects the behavior of the intrinsic flux (i.e., the background mentioned earlier) and has a minor impact on the extracted shift curves.
Taking ellipsoidal plus eclipsing binaries as an example, f}or $K > 30$~\kms, the typical precision in measuring $K$ and $e$ is within 10\% and 0.1, respectively.
We note that the random error in $K$ would increase with rising $e$ when $K$ is in the moderate range of 30 to 100~\kms.
The systematic offset, defined as the difference between the extracted and true values, is negligible.
For $K \le 30$~\kms, the photometric variations caused by the Doppler shift become small, making them difficult to detect in light curves obtained from ground-based telescopes.
Consequently, both random and systematic uncertainties become significant, making it challenging to accurately determine the orbital parameters in these regions.

\subsubsection{Shift Factor Precision in Realistic Observations}
\label{sec:SF}
Accurate determination of $S'$, i.e., the beaming factor $B$ for the $C$-band and the shift factor $S$ for the $A$-band, is crucial for deriving orbital parameters. In principle, this could be achieved if well-calibrated spectra of the targets were available. However, spectra from large-scale surveys are generally not well flux-calibrated. Alternatively, target spectra can be generated using their atmospheric parameters.
The latter now can be obtained either from massive spectroscopic surveys like LAMOST \citep{lamost_para,lamost_para2} or from  narrow/medium photometric surveys \citep{beyond_1,beyond_2,beyond_3}.

In this regard, the uncertainties in determining $S$ and $B$ can be assessed based on the uncertainties in atmospheric parameters, which are typically 150~K for $T_{\rm{eff}}$, 0.2~dex for log~$g$, and 0.2~dex in [Fe/H].
In addition to atmospheric parameters, $S$ is also influenced by the profile of strong absorption lines, which can be affected by the star's rotation speed ($v\sin i$). Since this parameter is difficult to measure, we assume it is uniformly distributed within empirical ranges for different types of stars \citep{rotate_speed,rotate_speed2} (see Figure \ref{fig:example} (c)). 
The uncertainties in $S$ and $B$ are then calculated using a Monte Carlo procedure. 
For each grid point, we generate 10,000 target spectra by sampling atmospheric parameters ($T_{\rm{eff}}$, log~$g$, and [Fe/H]) and $v\sin i$. The first three parameters are assumed to follow Gaussian distributions, while $v\sin i$ follows a uniform distribution.
The uncertainties (given by half of the 68\% interval) of $S$ and $B$ are determined by their distributions, calculated from 10,000 spectra using the method in \cite{Z24}.
The final results are presented in Figure \ref{fig:example} (d), where the typical uncertainty in $S'$ is approximately 5.5\%, leading to an error in the measured $K$ of a similar magnitude (see the red line at 6\% in the left panel of Figure \ref{fig:vary}).

\subsubsection{Overall Performance}
The final uncertainty\footnote{For eccentricity $e$ the error is dominated by the photometric measurements only.} of the extracted $K$ is again shown in Figure \ref{fig:vary}, by combining the errors from both photometric measurements and determinations of $S'$.
The plots clearly show that, at large  values ($K > 100$ \kms), the uncertainty is primarily dominated by the error in $S'$, while at small values ($K < 30$ \kms) it is dominated by the uncertainty in photometric measurements. In the intermediate range ($K$ between 30 and 100~\kms), both sources of uncertainty contribute to the overall error.

Our comprehensive simulations demonstrate that the ``Frog-eyes'' system can be effectively used to systematically search and measure orbital parameters for SB1 with $K \ge 30$~\kms\ in a photometric approach from ground-based observations.
Its ability to measure $K$ is comparable to that of low-resolution spectroscopy with $R \sim 800$, but with photometric efficiency. 
For SB1, let's consider one solar-type star ($\sim$1 M$_{\odot}$) as the primary  and an M-type star ($\sim$0.5 M$_{\odot}$) as the secondary star.
Given the above measurement capability for $K$, the upper limit of the orbital period is $P_{\rm max} =20$ days.
If the companion is a compact object (assuming 3 M$_{\odot}$), this upper limit increases to $P_{\rm max} = 606$ days.
By relaxing the detection threshold to 3$\sigma$ (${K}/{\sigma_K}=3$), $P_{\rm max}$ extends to 160 days for an M-type companion and 4845 days for a compact object.
By incorporating stellar population and evolution in further simulations, we could present detailed estimates of the number of interesting binary systems that can be detected using 1.5m wide field-of-view survey telescopes equipped with the $A$- and $C$-band systems.

\section{Validation with Real Observations}
\label{sec:validate}
\begin{figure*}
    \centering
    \includegraphics[width=\textwidth]{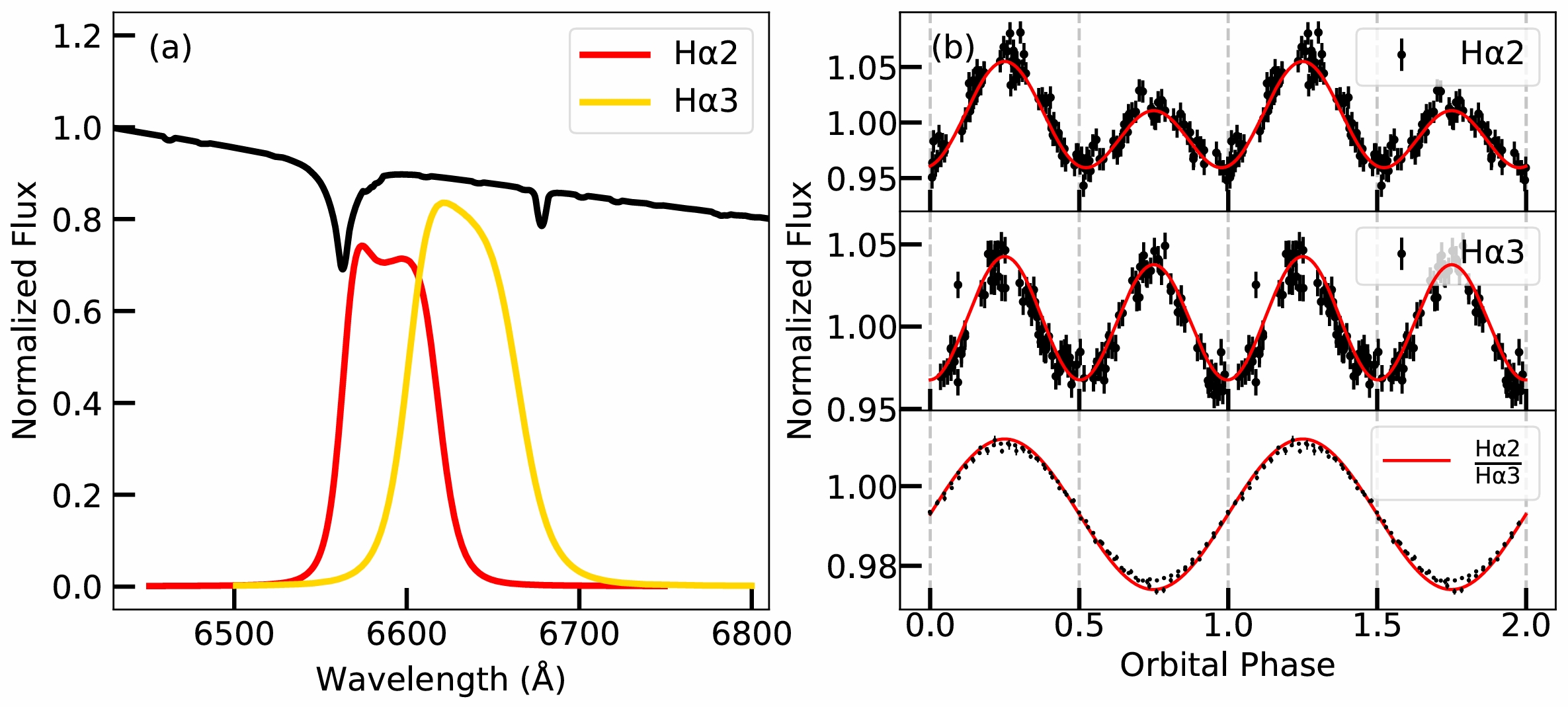}
    \caption{\textbf{Derived light curves of HD 265435 with the $\rm{H\alpha 2}$ and $\rm{H\alpha 3}$ filters. (a)} Normalized synthetic spectra of HD 265435 and transmission curves of $\rm{H\alpha 2}$ and $\rm{H\alpha 3}$. 
    \textbf{(b)} \zcj{The upper} two panels show normalized observed light curves (black dots) and our fitting results (red lines) for $\rm{H\alpha 2}$ and $\rm{H\alpha 3}$, respectively.  The bottom panel shows the predicted shift curve (black dots) from real RV measurements \cite{HD265435} and the ratio (red line) between the $\rm{H\alpha 2}$ and $\rm{H\alpha 3}$ fitting results.}
    \label{fig:HD265435}
\end{figure*}

In this section, we demonstrate the capability of our ``Frog-eyes'' system to measure orbital parameters for \zcj{SB1} systems using real observations.
To achieve this, we conducted test observations with the Chinese 2.16 m telescope at Xinglong Observatory on the bright star HD 265435 \citep{HD265435}, which has a magnitude of $V = 11.8$. 
It is a close ($P \sim 100$ min) hot subdwarf-white dwarf system with a spectroscopically determined $K$ of 343~\kms.
We selected the $\rm{H}\alpha\rm{2}$ and $\rm{H}\alpha\rm{3}$ filters (see their profiles in Figure \ref{fig:HD265435} (a)) installed in the BAO Faint Object Spectrograph and Camera (BFOSC) \cite{216}, which are part of a series of narrow-band $\rm{H}\alpha$ filters (with central wavelengths ranging from 656.2 to 706.0~nm) designed for observing star-forming regions in nearby galaxies at various red-shifts \cite{216,Ha_curve}.
Although the two filters were not specifically designed for our ``Frog-eyes'' system, $\rm{H}\alpha\rm{2}$ is very close to the strong $\rm{H}\alpha$ absorption line at 6563~\AA, making it suitable for use as the $A$-band. $\rm{H}\alpha\rm{3}$ primarily covers the continuum region and can thus be used as the $C$-band.
The overall shift factor $S'$ is approximately 15.2 when using synthetic spectra based on the atmospheric parameters and $v \sin i$ measured in \cite{HD265435}. Although this factor is about one-third of that obtained with our optimal filter design, it is still an order of magnitude larger than its beaming factor. As a result, it could cause flux variations at the level of $\sim$4 percent, which are detectable with a ground-based telescope.

On Dec. 9, 2023, HD~265435 was observed for 5 hours using the $\rm{H}\alpha\rm{2}$ and $\rm{H}\alpha\rm{3}$ filters installed \zcj{in the} BFOSC.
We adopted an observation strategy that employs alternative runs between the two filters.
For each run, five exposures were taken, each with an exposure time of 60 seconds.
In total, 115 and 118 exposures were performed in  $\rm{H}\alpha\rm{2}$ and $\rm{H}\alpha\rm{3}$, respectively.
The typical signal-to-noise ratio ranges from 300 to 500, depending on seeing and weather conditions\footnote{The observing conditions were suboptimal due to the presence of thin cloud cover.}.
These images were processed using a standard reduction procedure, which included bias subtraction, flat-field correction, and cosmic ray removal.
The differential aperture fluxes for HD 265435 were subsequently computed for each epoch in both two filters, using two nearby bright stars (with $G$ magnitudes of 12.3 and 13.6, respectively) as reference stars.
Using other relative bright field stars as checks, the typical  1$\sigma$ uncertainties are approximately 1\% and 0.8\% for $\rm{H}\alpha\rm{2}$ and $\rm{H}\alpha\rm{3}$, respectively.
The normalized light curves, obtained using mean fluxes, are presented in Figure \ref{fig:HD265435}\,(b).
The $\rm{H}\alpha\rm{3}$ results reveal a clear sinusoidal-like ellipsoidal light curve, caused by the tidal forces of its compact companion. 
This curve closely resembles the one observed by TESS. as reported by Pelisoli et al. \citep{HD265435}.
For $\rm{H}\alpha\rm{2}$, the effect of \zcj{the} Doppler shift is clearly evident in the light curve, as indicated by the significant differences between neighboring peaks.
This difference is approximately 4 percent, in excellent agreement with our expectations.

In principle, the orbital parameters can be directly extracted by fitting Equation \ref{eq:divide_f} to the shift curve $\frac{F_{\rm{H}\alpha\rm{2}}}{F_{\rm{H}\alpha\rm{3}}}$.
However, the light curves for these two filters are not obtained simultaneously.
As an alternative, we fit model light curves represented by Equations \ref{eq:ha2} and \ref{eq:ha3} to the real observational ones of $\rm{H}\alpha\rm{2}$ and $\rm{H}\alpha\rm{3}$, respectively.
In the case of HD 265435, the intrinsic flux variation, i.e., $F_{0,A}(t)$ and $F_{0,C}(t)$, is caused by the tidal forces of the compact object in the binary system. This sinusoidal-like ellipsoidal modulation can be easily modeled using a fifth-order Fourier series function with a fixed period reported in \cite{HD265435}.
We use the Markov Chain Monte Carlo (MCMC) technique for light curve modeling. A total of 12,000 steps were run, with the first 2,000 steps serving as the burn-in phase.
The posterior distributions of the orbital parameters yielded by the fitting are shown in Supplementary Figure S5.
From the distributions, the value of $K$ is well-constrained and estimated to be $377^{+26}_{-25}$\kms. This result is consistent with the spectroscopic measurement of $343.1 \pm 1.2$\kms reported by \citep{HD265435} under the assumption of a circular orbit, with a mild offset of less than 10\%.
Repeating the analysis under the assumption of a circular orbit ($e = 0$), as in \citep{HD265435}, yields a $K$ value of $364 \pm 28$~\kms, effectively reducing the offset by nearly half.
The random uncertainty of $K$ is mainly contributed from the 1-2\% photometric uncertainty.
The slight offset in $K$ between spectroscopy and photometry may result from: (1) discrepancies between the manufacturer-provided transmission curves and the actual ones for $\rm{H}\alpha\rm{2}$ and $\rm{H}\alpha\rm{3}$, and (2) systematic uncertainties in the stellar parameters used to generate synthetic spectra for HD~265435.
This offset could be further reduced with accurate measurements of the filter transmission curves or by directly extracting $S'$ from high-precision observed spectra of HD~265435.
The eccentricity $e$ is weakly constrained, with its posterior distribution resembling that of a circular orbit. The profile exhibits a monotonic increase as $e$ decreases, indicating a preference for small values. Consequently, an upper limit of 0.14 is adopted as the final estimate for $e$.

These results clearly demonstrate that the ``Frog-eyes'' system can indeed measure the orbital parameters of binaries photometrically. Even with non-specialized filters and modest photometric precision, the expected  $\sim$4\% modulation on flux caused by the Doppler shift is clearly detected, and the extracted orbital parameters are consistent with those obtained through traditional spectroscopy. In future work, we expect to reduce both random and systematic uncertainties by using an optimal design for this ``Frog-eyes'' system, and employing more advanced data analysis techniques.

\section{Caveats and Challenges}
In the earlier sections, we have demonstrated that the ``Frog-eyes'' system offers both broad applicability for identifying binary systems and high precision in extracting orbital parameters for SB1 systems, even with its simple trapezoidal design and ground-based photometric precision. While these preliminary results are promising, this innovative approach still faces several caveats and challenges that need to be addressed.

1) The ``Frog-eyes'' system is highly efficient for systems exhibiting single-lined periodic motions and performs particularly well for SB1 systems, including the precise measurements of orbital parameters. For double-lined binaries, significant contributions from the anti-phase RV motions of the secondary component can result in anti-phase flux modulations in the $A$-band, and reduce the overall signal strength. Although the ``Frog-eyes'' system remains applicable in such cases, determining orbital parameters becomes challenging, requiring extremely high photometric precision. 
Fortunately, among the millions of targets observed by LAMOST, only about 1\% are classified as double-lined or triple-lined spectroscopic binaries \citep{sb2_fing_1,sb2_find_0}. However, we note that this fraction represents a lower limit, as these studies primarily focus on detecting significant SB2 systems without consideration of systems with  $K\le$ 30 \kms. We will conduct a detailed binary population synthesis to estimate the number of single-lined and double-lined spectroscopic binaries that can be detected using the ``Frog-eyes'' system.

2) Accurate conversion of shift fluxes into RV variations requires stable absorption line profiles (i.e., a constant $S$). Systems such as stars with stellar spots (e.g., see Figure 2 of \cite{spot_doppler}) or binaries with accretion may exhibit variations in their line profiles, which prevent the ``Frog-eyes'' system from accurately extracting orbital parameters of SB1 systems.
While these variations can be precisely modeled through a combination of light curve modeling and synthetic spectra, this approach is too complex for large surveys. Currently, there is no straightforward solution or specialized filter design to account for these potential line profile variations when measuring orbital parameters. However, at the very least, the ``Frog-eyes'' system is effective at identifying these rapid line profile variations, which, in some cases, are more intriguing than the binary systems themselves. In the future, by leveraging machine learning techniques, we may be able to resolve this degeneracy, detecting these line profile variations and accurately measuring orbital parameters simultaneously.

\zcj{3) It is challenging to cover all target types with a single set of ``Frog-eyes'' filters that focus on just one absorption line, as the line profile evolves significantly with changes in effective temperature. For example, the H$_{\rm \delta}$ line becomes weak and even disappears in M dwarfs, which is why the design described in Section 3 is not suitable for targets with effective temperatures below 6000~K. To overcome this limitation, incorporating multiple sets of filters locating different absorption lines or even molecular lines (for example, see Supplementary Section B and Supplementary Figure S2) is a feasible solution.}

\zcj{4) As a novel method, the ``Frog-eyes'' system imposes new requirements on the industrial production of both filters and telescopes. For example, the filters must be sufficiently large to achieve a wide field-of-view while maintaining uniform response across the focal plane. These stringent requirements present significant challenges in manufacturing and quality control. To address these challenges, we have made preliminary progress through collaborations among various institutions, including but not limited to the National Astronomical Observatories of China (NAOC), the Nanjing Institute of Astronomical Optics and Technology, and Zhejiang University. Our teams have secured funding to establish an optical laboratory at NAOC, focused on high-precision measurements of filter transmission curves across the focal plane. This initiative is intended to ensure that the filters meet the stringent standards required by the ``Frog-eyes'' system.}

\zcj{In summary, as a newly proposed method, the ``Frog-eyes'' system has a long path to go before it can be considered mature or even classical. However, we are progressively advancing the project through a series of strategic initiatives. We are confident that, in the near future, the ``Frog-eyes'' system will provide radial velocity curves for hundreds of millions targets, driving significant advancements across various fields.}

\section{Summary}
To bridge the significant gap between the rapid detection of variable stars and the slow accumulation of RV curves, this work proposes a new method called ``Frog-eyes'', capable of photometrically measuring \zcj{RV variations, particularly for SB1 systems,} at a rate orders of magnitude faster than traditional spectroscopy.
The ``Frog-eyes'' system employs a pair of narrow-band filters: the $A$-band, positioned near a strong absorption line, and the $C$-band, which focuses on the nearby continuum.
Any variation in RV, \zcj{even those originating from radial pulsations, will} cause a corresponding change in the flux difference between the two filters. This flux variation is at least an order of magnitude greater than Doppler beaming, and can be detected even with ground-based telescopes.

With an optimized design based on current manufacturing technology, this method is widely applicable to stars with $T_{\rm eff} \geq 6000$\,K. Under typical ground-based photometric precision (0.3\%), mock data simulations demonstrate that it achieves high precision in orbital parameter measurements, with uncertainties below 10\% for $K$ and 0.1 for $e$, in binary systems with $K \geq 30$ \kms.
Improvements in filter design or precision of both photometry and atmospheric parameter measurement could further boost the method's performance, expanding its applicable range and increasing its precision.
The capability of this system is further validated by real observations of the binary system HD 265435. The detection of the predicted $\sim 4$\% flux modulation and the excellent agreement between the extracted photometric $K$ and the spectroscopic value reported in \cite{HD265435} effectively demonstrate the method's applicability and the robustness of the theoretical simulations.

Finally, we emphasize the significant potential of this method in accelerating the acquisition of RV curves through cost-effective photometric observations. Even small ground-based telescopes, such as those with 50-cm apertures, can utilize this approach to economically probe RV variations for all-sky bright sources. Looking ahead, we plan to implement this method in next-generation time-domain surveys like SiTian. 
This will enable a dramatic increase in the efficiency of binary detection and orbital measurements, surpassing spectroscopic techniques by several orders of magnitude. 
We encourage broader collaborations in this promising field to drive a transformative breakthrough in the time-domain arena, revealing the fundamental physics that govern our dynamic Universe.

\vspace{0.6cm}

{\footnotesize
We are grateful to Professor Haibo Yuan for his valuable suggestions on filter design and further discussions.

This work acknowledges support from the National Natural Science Foundation of China (NSFC) for grants Nos. 12422303,  11988101, 11933004, and 12273057. YH acknowledges support from the National Key Basic R\&D Program of China via 2023YFA1608303 and the China Manned Space Project with No. CMS-CSST-2021-A08. JFL acknowledges support from the New Cornerstone Science Foundation through the New Cornerstone Investigator Program, the XPLORER PRIZE, and the International partnership program of the Chinese academy of sciences (Grant No. 178GJHZ2022047GC).
SW acknowledges support from the National Key Basic R\&D Program of China via 2023YFA1607901.
T.C.B. acknowledges partial support for this work from grant PHY 14-30152; Physics Frontier Center/JINA Center for the Evolution of the Elements (JINA-CEE), and OISE-1927130: The International Research Network for Nuclear Astrophysics (IReNA), awarded by the US National Science Foundation.

We acknowledge the support of the staff of the  Xinglong 2.16m telescope. This work was partially supported by the  Open  Project  Program  of  the  Key  Laboratory  of  Optical  Astronomy,  National Astronomical Observatories, Chinese Academy of Sciences. 
This research made use of {\tt Photutils}, an Astropy package for
detection and photometry of astronomical sources, which is publicly available at \url{https://photutils.readthedocs.io/en/stable/index.html}.

\vspace*{0.6cm}

{\bf Conflict of interest} The authors declare that they have no conflict of interest.

\bibliographystyle{scibull}
\bibliography{refer}
}

\newpage
\begin{appendix}




\renewcommand{\thesection}{Supplementary Materials}
\renewcommand{\thesubsection}{\Alph{subsection}}
\newpage
\setcounter{table}{0}   
\setcounter{figure}{0}
\renewcommand{\thetable}{S\arabic{table}}
\renewcommand{\thefigure}{S\arabic{figure}}
\section{}
\subsection{Filter Design}
\label{filter_design}
As discussed in the main text, the filter design takes into account two critical factors: the main body, which includes the central wavelength and effective width, and the profile of the `smooth area'. The two factors collectively affect the filter's applicability across various stellar spectral types and radial velocity (RV) ranges, its effectiveness in amplifying RV signals, the observational limiting magnitude, and the complexity of industrial production. Although these factors are interrelated, our goal is to assess their individual impacts and  achieve a balanced design for optimal performance.

The influence of these factors can be summarized as follows. First, for the main body, the central wavelength determines which strong absorption line is monitored for red-shift or blue-shift. 
Absorption lines with larger equivalent widths are preferred, as they exhibit greater flux variations in response to RV changes.
Moreover, the absence of absorption features near this strong absorption line is essential for more accurate monitoring of the continuum with the $C$-band.
Secondly, a narrower effective width generally results in a larger shift factor, but a shallower limiting magnitude and greater difficulty in industrial production, whereas a wider width tends to have the opposite effects.
Finally, the profile of the `smooth area' affects the ``Frog-eyes'' system's ability to smooth and amplify RV signals, while also introducing challenges for industrial production.
Ideally, the profile should allow the absorption line center to move within the `smooth area' to maximize flux variation. 
A `smooth area' wider than the line wings can effectively smooth the RV-modulated flux but reduces the amplitude of variation, whereas a narrower width increases the amplitude but is less effective at smoothing.

In this work, these factors are determined as follows. 1) The prominent H$\delta$ absorption line ($\sim$4102~\AA) is selected for our simulation. The performance of other lines, such as H$\alpha$, will be investigated in future work dedicated to filter design. 2) Two profiles of the `smooth area' are evaluated: a linear gradient and a Gaussian-like transition, resulting in either a trapezoidal or Gaussian filter shape, respectively. The trapezoidal profile is finally selected because it offers better smoothing and amplification of RV signals under the same effective width. In this profile, the `smooth area' refers to the right-angled triangular region at the shorter-wavelength end of the filter. 3) The width of the `smooth area', defined as the distance between the left and right bottom edges of the right-angled triangle, is set to 10~\AA, with the left edge at a wavelength of 4097.5~\AA. This configuration balances a wider linear response range with a higher shift factor, as determined by a grid search\footnote{The grid of widths tested includes 1, 2, 5, 10, and 15~\AA. The grid of the left edge wavelength ranges from 4082.5~\AA, to 4102.5~\AA, with a step of 1~\AA.}. 4) The effective width is chosen to be 50~\AA~ (see Supplementary Tables \ref{tab:60s} and \ref{tab:180s} for additional options) to balance the shift factor, limiting magnitude, and challenges of industrial production. The central wavelength of the filter is then determined to be 4127.5~\AA.

\subsection{Alternative Filter Designs}
\label{alternative_design}
\zcj{By positioning the $C$-band at the blue end of an absorption line (see $A_2$-band in Supplementary Figure \ref{fig:NEW_C}), the obtained light curve would contain a similar but anti-phase RV-modulated signal to that of the $A$-band. Compared to the design described in Section 3, this configuration can effectively double the RV modulated signal while slightly reduce the precision in monitoring other intrinsic flux variations.}

\zcj{By placing the narrow-band filter pairs near strong molecular bands (see Supplementary Figure \ref{fig:M3700}), the obtained shift flux can also exhibit a linear response to the RV. This means the ``Frog-eyes'' can be applicable to different types of targets by modifying the filter designs.}

\section*{Author Contribution}
C.J.Z. proposed the method, inspired by discussions led by J.F.L. and Y.J.L. and regular discussions organized by Y.H., and was responsible for filter design, data analysis, and preparing the initial draft. Y.H. and J.F.L. led the project, guiding filter design and mock data simulations, and finalized the draft. Y.H. also coordinated the observation time with the 2.16m telescope.
H.W. joined discussions, supplied guiding suggestions on observation, and data analysis.
T.C.B. joined discussions and modified the manuscript.
Other authors contributed in joining discussions and supplied useful suggestions to this work.

\begin{table*}[h]
    \centering
    \caption{Performance of different filter designs with \zcj{a} single 60-second exposure.}
    \begin{threeparttable}
    \begin{tabular}{cccccccc}
    \hline
    \hline
       Filter $^a$& $W_{\rm eff}$ (\AA)&$D_{2} \le 10\%$ $^b$ &$S_{\rm{median}}$ $^b$&$\sigma_S$ $^b$&$m_{\rm S/N = 3}$ $^c$&$m_{\rm S/N = 10}$ $^c$&$m_{\rm S/N = 100}$ $^c$\\
       \hline
       $A_1$&10&0.00\%&n.a.&n.a.&19.17&17.40&12.76\\
       $A_2$&20&87.43\%&125.43&6.29\%&19.89&18.13&13.51\\
       $A_3$&30&88.01\%&82.60&6.55\%&20.30&18.56&13.95\\
       $A_4$&40&88.49\%&63.25&5.72\%&20.58&18.85&14.26\\
       \rowcolor{red!30}$A_5$&50&88.84\%&49.49&5.75\%&20.80&19.08&14.50\\
       $A_6$&60&82.05\%&42.31&5.38\%&20.97&19.26&14.70\\
       $A_7$&65&89.07\%&38.81&5.56\%&21.05&19.34&14.79\\
       $A_8$&140&92.37\%&18.67&4.82\%&21.73&20.08&15.62\\
       \hline
    \end{tabular}
         \begin{tablenotes}
        \footnotesize
        \item[$a$] The highlighted $A_5$ filter is the one shown in Figure \ref{fig:example} (a). Other listed filters share the same trapezoidal profile and `smooth area', but have modified main bodies to achieve different effective widths.
        \item[$b$] The column $D_{2}\le 10\%$ refers to the fraction of grid points with $D_{2}\le 10\%$. The full grid is defined by 5000\,K $\le T_{\rm eff} \le 12,000$\,K with a step of 200\,K, $2.0 \le$ log\,$g\le 5.0$ with a step of 0.1 dex, and $v$sin$i$ at discrete points: 10, 50, 70, 100, 150, 200, 250, 300, and 400 \kms. $S$ and $\sigma_S$ refer to the median value and uncertainties of the shift factor, respectively. All three columns are computed for solar abundance with {\tt PHOENIX} \citep{PHOENIX} synthetic spectra. 
        \item[$c$] The limiting magnitudes at different signal to noise (S/N) ratios. Observations are assumed to be performed with a 1-meter telescope at Lenghu observatory \citep{lenghu}.
    \end{tablenotes}
    \end{threeparttable}
    \label{tab:60s}
\end{table*}

\begin{table*}[h]
    \centering
    \caption{Performance of different filter designs with three continuous 60-second exposures.}
    \begin{tabular}{cccccccc}
    \hline
    \hline
       Filter & $W_{\rm eff}$ (\AA)&$D_{2} \le 10\%$ &$S_{\rm{median}}$&$\sigma_S$ &$m_{\rm S/N = 3}$ &$m_{\rm S/N = 10}$ &$m_{\rm S/N = 100}$ \\
       \hline
       $A_1$&10&0.00\%&n.a.&n.a.&19.86&18.27&13.94\\
       $A_2$&20&87.43\%&125.43&6.29\%&20.57&18.99&14.69\\
       $A_3$&30&88.01\%&82.60&6.55\%&20.98&19.41&15.13\\
       $A_4$&40&88.49\%&63.25&5.72\%&21.26&19.70&15.44\\
       \rowcolor{red!30}$A_5$&50&88.84\%&49.49&5.75\%&21.48&19.92&15.68\\
       $A_6$&60&82.05\%&42.31&5.38\%&21.65&20.10&15.88\\
       $A_7$&65&89.07\%&38.81&5.56\%&21.72&20.17&15.96\\
       $A_8$&140&92.37\%&18.67&4.82\%&22.39&20.88&16.79\\
       \hline
    \end{tabular}
    \label{tab:180s}
\end{table*}
\newpage
\subsection{Supplementary Figures}
\begin{figure}[H]
    \centering
    \includegraphics[width=0.95\linewidth]{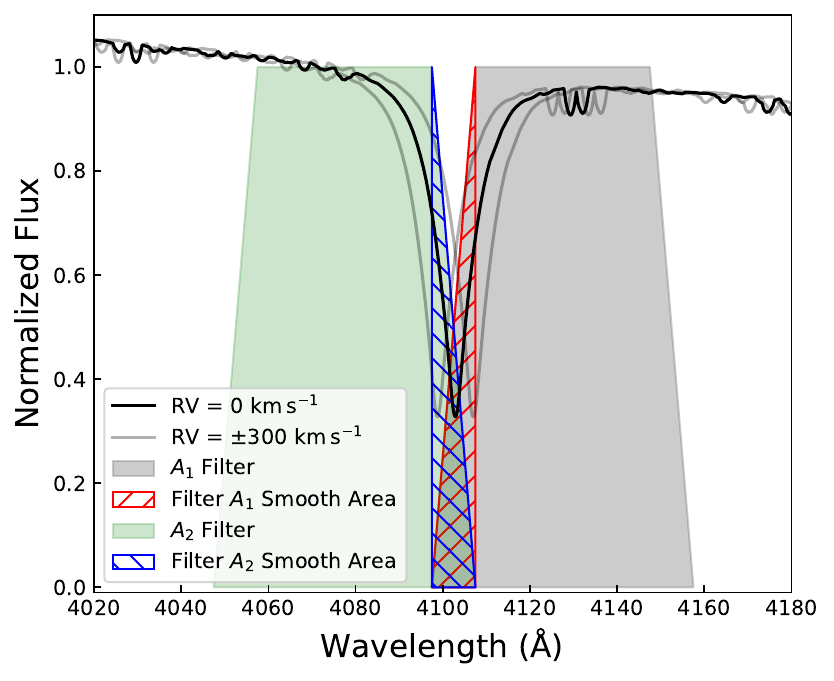}
    \caption{Transmission curves of an alternative design \zcj{for the} $A_1$- and $A_2$-band.}
    \label{fig:NEW_C}
\end{figure}

\begin{figure}[H]
    \includegraphics[width=0.95\linewidth]{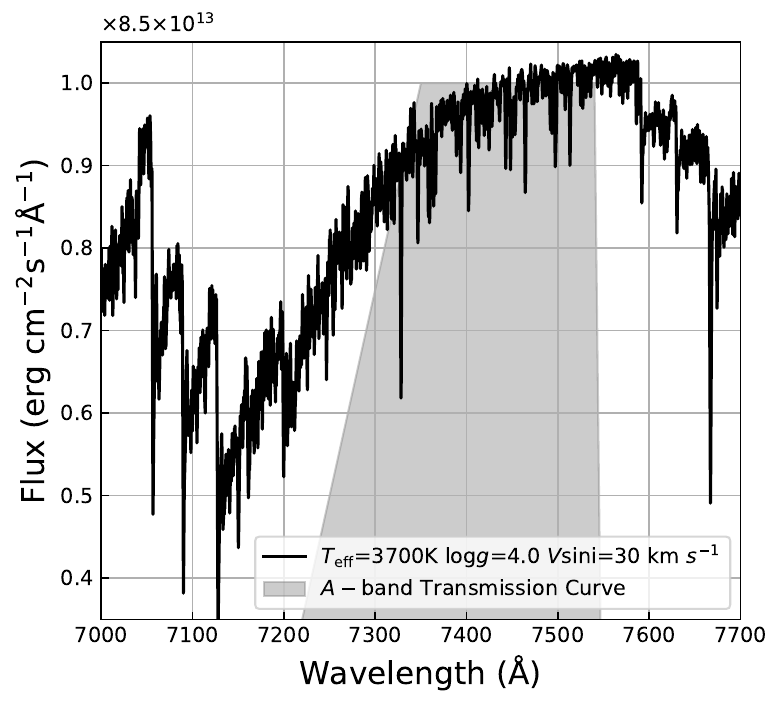}
    \caption{Similar to Figure \ref{fig:NEW_C}, but for a different design of the $A$-band, focusing on the molecular features of an M dwarf with an effective temperature of 3700\,K. The spectrum is degraded to a resolution of $R \sim 2000$ for visual clarity.}
    \label{fig:M3700}
\end{figure}
\begin{figure}[H]
    \centering
    \includegraphics[width=.45\textwidth]{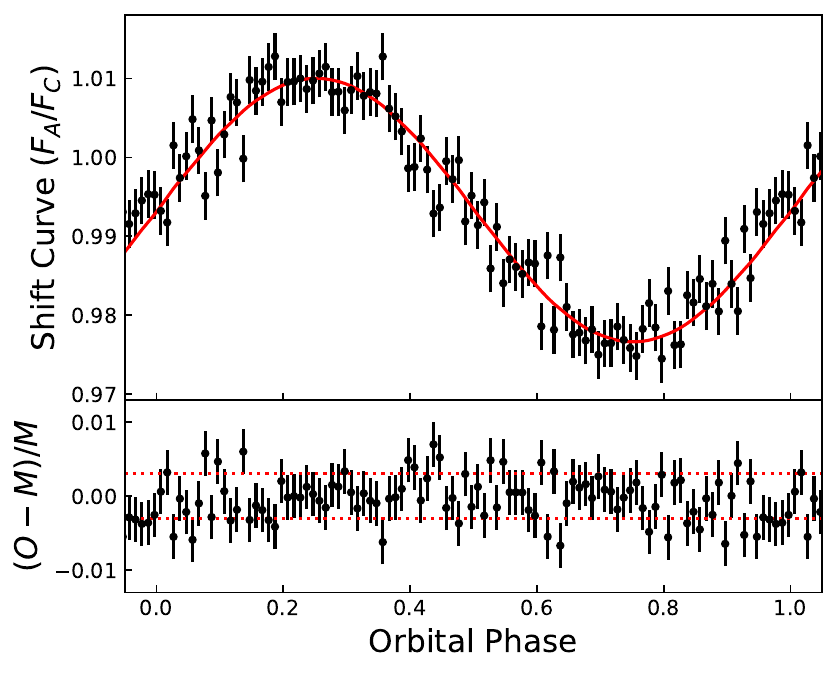}
    \caption{\textbf{Simulated shift curve for a system with \zcj{$K = 100$\,\kms.}} The upper panel shows the shift curve ($\frac{F_A}{F_C}$) of the system with the shift factor being 50.2 for the $A$-band.
    The beaming factor is 2.13 for the $C$-band. The lower panel shows the residuals between the simulated ($O$) and modeled ($M$) shift curve. Red dotted lines mark the $1\sigma$ scatter of the residuals, which is 0.31\%.}
    \label{fig:theory_curve}
\end{figure}

\begin{figure}[H]
    \centering
    \includegraphics[width=0.99\linewidth]{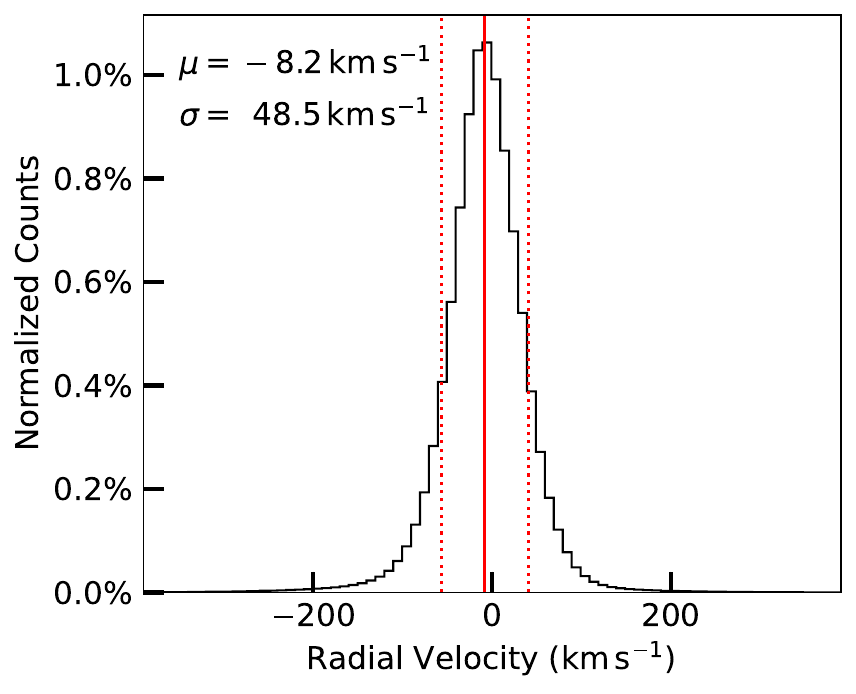}
    \caption{\textbf{Radial velocity distribution of field stars in the Milky Way collected by \zcj{the} LAMOST survey.} Data are from \zcj{the} LAMOST \citep{LAMOST1,LAMOST2,LAMOST3} DR11 catalog for A, F, G, and K stars. The median (indicated by \zcj{the} solid red line) and standard deviation (marked by \zcj{the} dashed red line) of the radial velocity distribution are -8.2~\kms and 48.5~\kms, respectively.}
    \label{fig:lamost}
\end{figure}
\begin{figure}[H]
    \centering
    \includegraphics[width=0.90\linewidth]{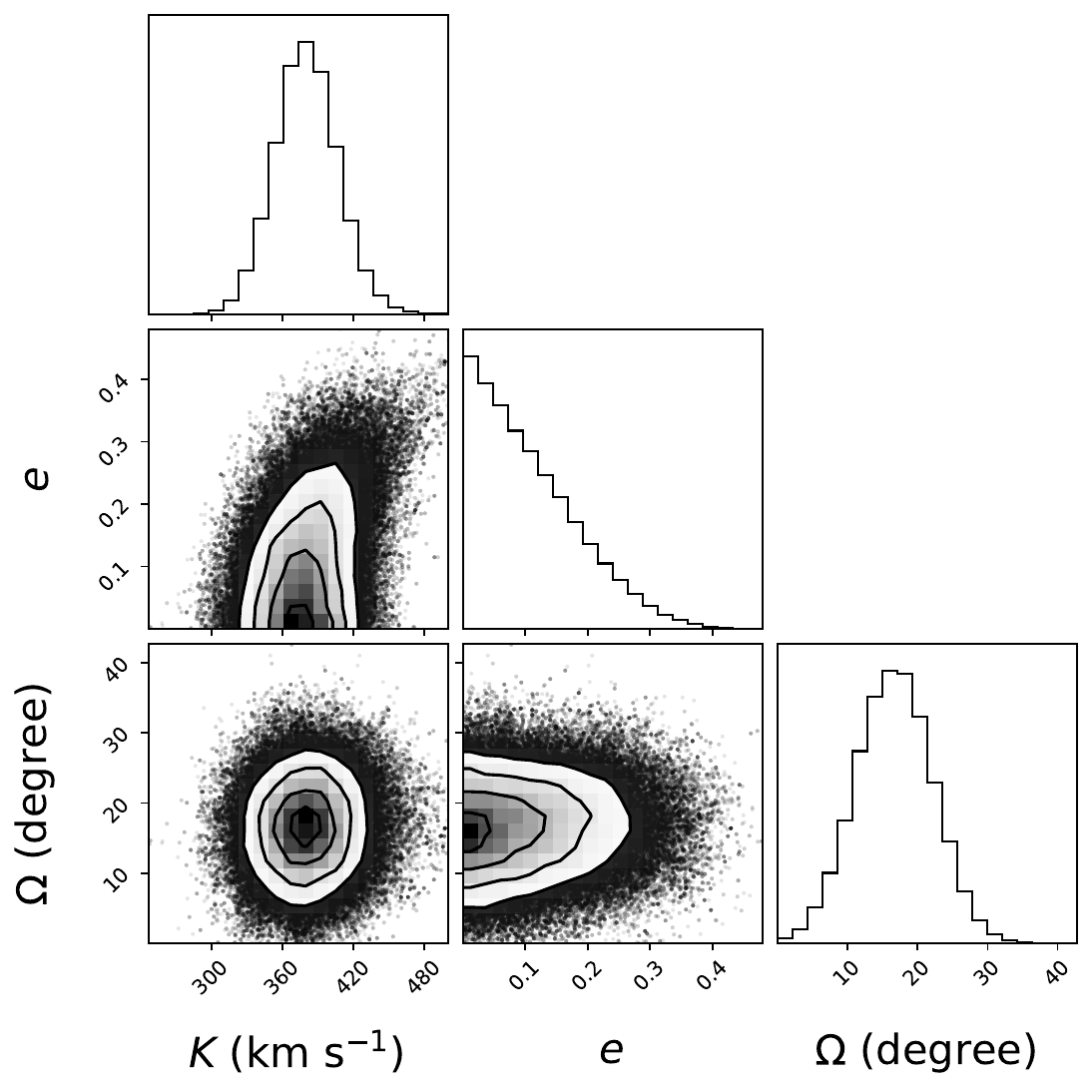}
    \caption{\textbf{Corner plot of the MCMC fitting results.} $K$, $e$, and $\Omega$ refer to semi-amplitude radial velocity, eccentricity, and argument of pariastron, respectively.}
    \label{fig:corner}
\end{figure}

\begin{figure*}
    \centering
    \includegraphics[width=0.32\linewidth]{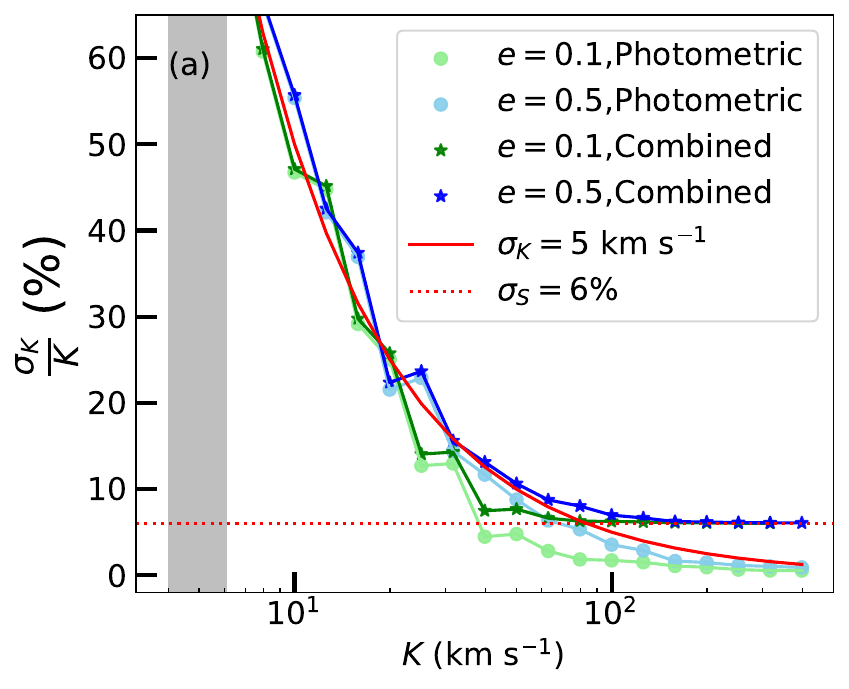}
    \includegraphics[width=0.32\linewidth]{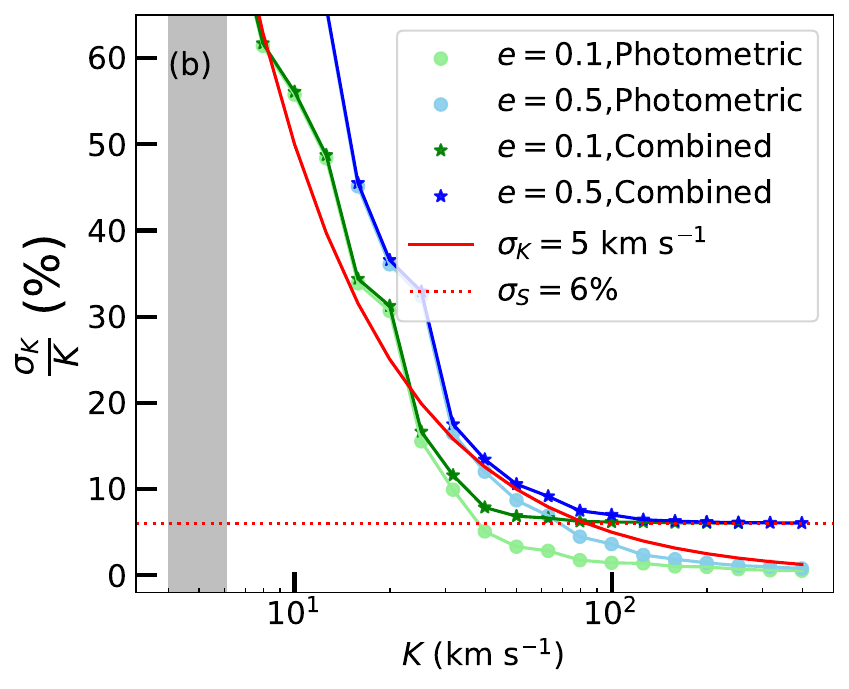}
    \includegraphics[width=0.32\linewidth]{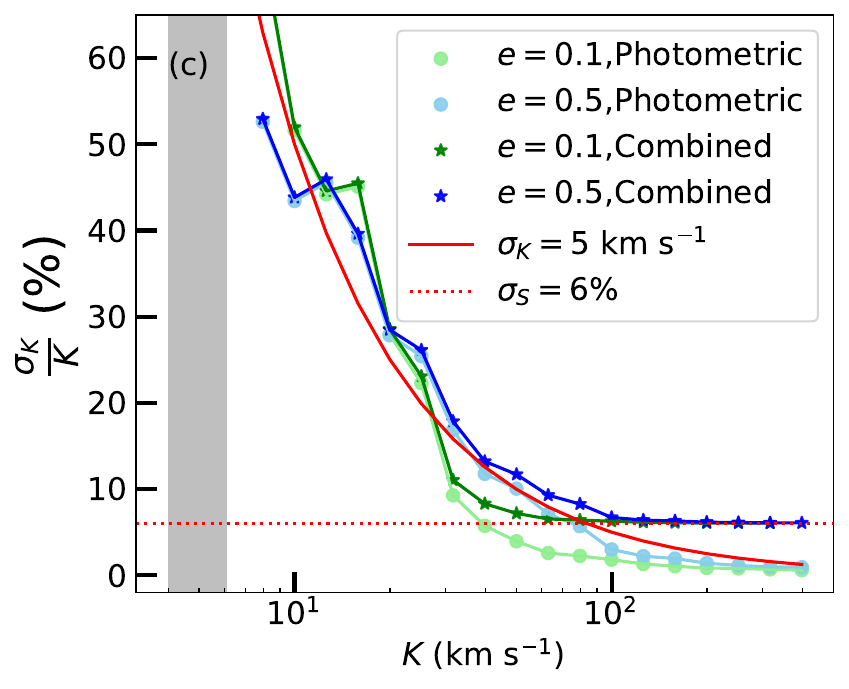}
    \caption{\textbf{Uncertainties of \zcj{the} measured $K$}. Similar to Figure \ref{fig:vary} (a), but for different \zcj{types of SB1 systems}. From left to right, the panels display the results for SB1 binaries without flux variation, with eclipses, and with ellipsoidal modulations, respectively.}
    \label{fig:K_A}
\end{figure*}

\begin{figure*}
    \centering
    \includegraphics[width=0.32\linewidth]{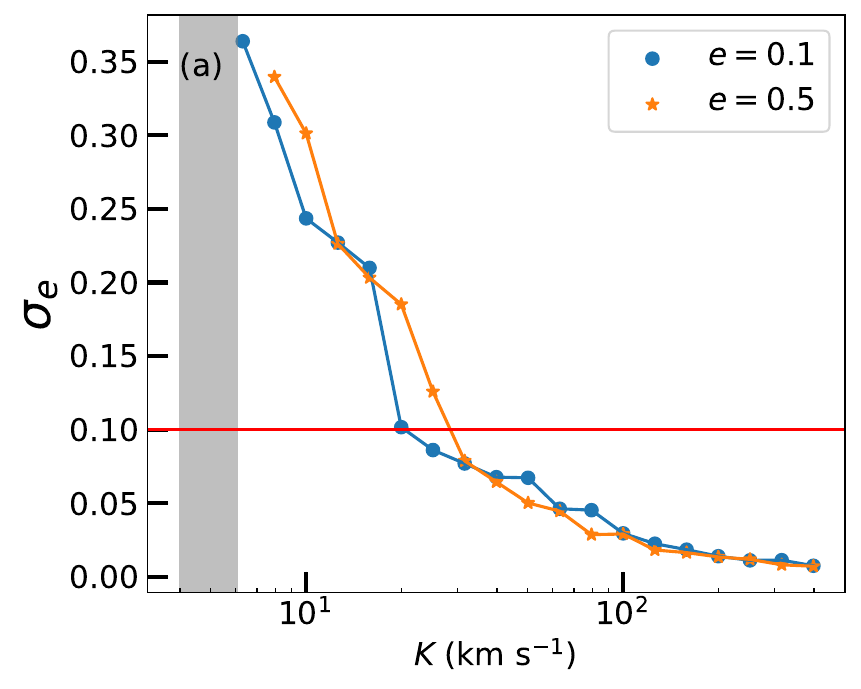}
    \includegraphics[width=0.32\linewidth]{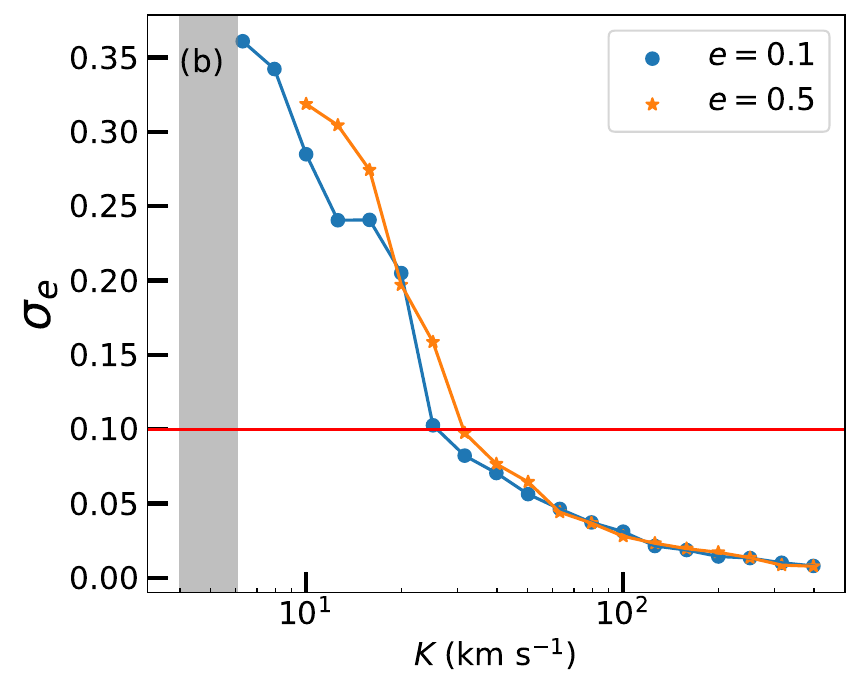}
    \includegraphics[width=0.32\linewidth]{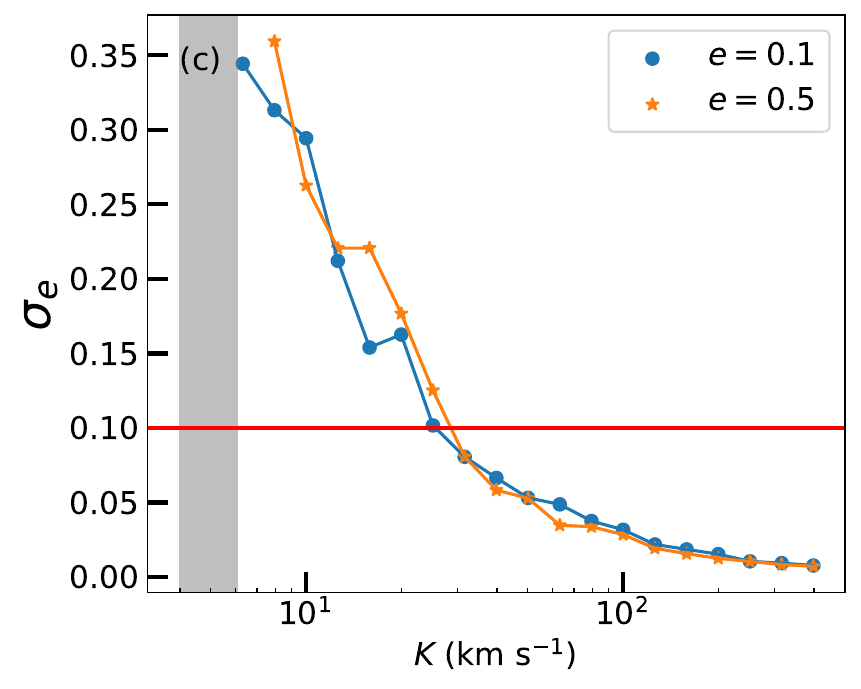}
    \caption{\textbf{Uncertainties of \zcj{the} measured $e$}. Similar to Figure \ref{fig:vary} (b), but for different \zcj{types of SB1 systems}. From left to right, the panels display the results for SB1 binaries without flux variation, with eclipses, and with ellipsoidal modulations, respectively.}
    \label{fig:E_A}
\end{figure*}
\end{appendix}
\end{multicols}
\end{document}